\documentclass[epj]{svjour}
\usepackage{epsf}
\begin{document}

\title{From independent particle towards collective motion in  
two electron square lattices}
\titlerunning{Independent particle and collective motion}

\author{Mois\'es Mart\'\i nez \and Jean-Louis Pichard}

\authorrunning{M. Mart\'\i nez and J.-L. Pichard}
        
\institute{CEA/DSM, Service de Physique de l'\'Etat Condens\'e, 
Centre d'\'Etudes de Saclay, 91191, Gif sur Yvette cedex, France}

\date{february 2002}

\abstract
{The two dimensional crossover from independent particle towards 
collective motion is studied using $2$ spinless fermions interacting 
via a $U/r$ Coulomb repulsion in a $L \times L$ square lattice with 
periodic boundary conditions and nearest neighbor hopping $t$. 
Three regimes characterize the ground state when $U/t$ increases. 
Firstly, when the fluctuation $\Delta r$ of the spacing $r$ between the 
two particles is larger than the lattice spacing $a$, there is 
a scaling length $L_0 = {\sqrt 8} \pi^2 (t/U)$ such that 
the relative fluctuation $\Delta r /\langle r\rangle$ is a universal function of
the dimensionless ratio $L/L_0$, up to finite size corrections of 
order $L^{-2}$. $L<L_0$ and $L>L_0$ are respectively the limits of 
the free particle Fermi motion and of the correlated motion of a 
Wigner molecule. Secondly, when $U/t$ exceeds a threshold $U^*(L)/t$, 
$\Delta r$ becomes smaller than $a$, giving rise to a correlated 
lattice regime where the previous scaling breaks down and analytical 
expansions in powers of $t/U$ become valid. A weak random potential 
reduces the scaling length and favors the correlated motion. 
} 


\maketitle

\section{Introduction}
\label{sec:intro}

 An important issue in quantum many body theory is to know how 
one goes from independent particle motion towards collective 
motion when one decreases the density $n_s$ of electrons repelling 
each other via a $U/r$ Coulomb repulsion. One can trap the electrons 
by a positive background of charges (jellium model) unless one 
uses periodic boundary conditions ($2d$ torus geometry). Many studies 
\cite{Egger,yannouleas1,yannouleas2,yannouleas3,peeters,hawrylak,filinov} 
assume a parabolic trap, its rotational invariance allowing to decouple the 
motion of the center of mass from the relative motions. A parabolic 
confinement has the merit to be realizable 
using semiconductor heterostructures \cite{hawrylak,dot} for electrons 
or electromagnetic fields for cold ions. It has however the 
disadvantage to yield a non uniform charge density, so 
that the formation of the electron solid when the confinement becomes 
weak results from a complicated interplay between edge and bulk 
orderings \cite{filinov,schweigert}. This leads us to study a system 
without edge, having the geometry of a two dimensional torus, where 
the density is uniform. For a continuous torus, the size of the corresponding 
Hilbert space is infinite, making the use of a truncated basis 
unavoidable for a numerical study. A way to avoid such a truncation 
is to take a lattice model (tight binding approximation). Considering the 
the simplest limit: the ground state (GS) of two spinless fermions 
in a square $L \times L$ lattice with periodic boundary conditions (BCs), 
we give an answer to four questions. Is there a simple one parameter 
scaling theory of the Fermi-Wigner crossover? How important are the finite 
size corrections to this scaling theory? Does the lattice play an important 
role? How those answers are modified when a weak random potential is included? 

The answers can be summarized as follows. In the strong coupling limit, 
the Coulomb repulsion pushes the spacing $|\vec{r}|$ between the two 
particle to take its maximum value $L/{\sqrt 2}$ and its fluctuations 
$\Delta r$ become of the order of the lattice spacing $a$. This gives 
rise to a strongly correlated lattice regime where analytical expansions 
in powers of $t/U$ are sufficient to describe the system. For weaker 
couplings, the relative motion becomes more and more delocalized and 
the lattice effects become irrelevant. This gives rise to a universal regime 
characterized by a scaling length $L_0 = {\sqrt 8} \pi^2 (t/U)$. This 
length corresponds to the system size for which the extra energy to add a 
second particle in the system is the same in the Wigner limit ($t=0$) 
than in the Fermi limit ($U=0$). $L<L_0$ is the Fermi limit, $L>L_0$ is 
the Wigner limit and the Fermi-Wigner crossover takes place when $L=L_0$. 
In this universal regime, the relative fluctuations $\Delta r/\langle r\rangle =
F(L/L_0)$ up to finite size corrections of order $L^{-2}$. When the site 
potentials have weak random fluctuations of order $W$, there is still a 
regime where one parameter scaling remains valid, $\Delta r/\langle r\rangle$ being
characterized by a function $F_W(L/L_W)$ which depends on $W$. Both  
$\Delta r/\langle r\rangle$ and the scaling length $L_W$ become smaller, in agreement
with the general idea that a weak disorder favors the correlated motion. 

\section{Lattice model}
\label{sec:hamiltonian}

We consider two electrons with symmetric spin wave functions and 
antisymmetric orbital wave functions (spinless fermions), free to move 
in a $L \times L$ lattice with periodic BCs, and interacting via a 
$U/|\vec{r}|$ repulsion. The Hamiltonian reads
\begin{equation}
H = -t \sum_{<i,j>} (c_i^{\dagger} c_j + h.c.) + \sum_i v_i n_i + 
\frac{U}{2} \sum_{i\neq j} \frac{n_i n_j}{|\vec{r}_{ij}|}
\label{siteH}
\end{equation}
where $i$, $j$ label the lattice sites, $<i,j>$ means $i$ nearest 
neighbor to $j$, $c_i^{\dagger}$, $c_i$ are the creation, annihilation 
operators of a spinless fermion at the site $i$; $n_i=c_i^{\dagger}c_i$ 
is the occupation number at the site labelled by the vector 
$\vec{i}=(i_x,i_y)$. The vector $\vec{r}_{ij}=\vec{i}-\vec{j}$ is
defined as the shortest vector going from the site $i$ to the site $j$ 
in a square lattice with periodic BCs. This means that $r_{x} \leq L/2$ 
and $r_y \leq L/2$ if $L$ is even, $L \rightarrow L-1$ if $L$ is odd. 
Hence, the pairwise interaction $U/|\vec{r}|$ exhibits a singularity on 
the lines $r_{x}=L/2$ and $r_y=L/2$ and another one at their crossing 
point $r_x=r_y=L/2$. $t=\hbar^2/(2m a^2)$ is the hopping term, $a$ the 
lattice spacing, $v_i$ the site potentials which are randomly distributed 
in the interval $[-W/2, W/2]$ and $U=e^2/(\epsilon a)$ the Coulomb 
interaction between two fermions separated by $a$ in a medium of 
dielectric constant $\epsilon$. When there is no disorder ($W=0$), 
$\vec{k} =(k_x,k_y)$ being the one particle momentum, it is more 
convenient to write $H$ using the Fourier transforms of the creation 
and annihilation operators. One has the relations
\begin{equation}\label{fouriercj}
c_{\vec{j}} = \frac{1}{L} \sum_{\vec{k}} a_{\vec{k}} e^{i \vec{k}\cdot 
\vec{j}},
\end{equation}
and
\begin{equation}
a_{\vec{k}} = \frac{1}{L} \sum_{\vec{j}} c_{\vec{j}} e^{-i\vec{k} \cdot 
\vec {j}}. 
\end{equation}
which yield
\begin{equation}\label{Hk}
H = \sum_{\vec{k}} a_{\vec{k}}^{\dagger} a_{\vec{k}} \, \varepsilon(\vec{k}) 
+\sum_{\vec{q},\vec{k}_1,\vec{k}_2} a_{\vec{k}_2+\vec{q}}^{\dagger}
a_{\vec{k}_2} a_{\vec{k}_1-\vec{q}}^{\dagger} a_{\vec{k}_1} V(\vec{q})
\end{equation}
where 
\begin{equation}\label{states1p}
\varepsilon (\vec{k}) = -2t \left( \cos k_x + \cos k_y \right)
\end{equation}
and
\begin{equation}
V(\vec{q}) = \frac{U}{2L^2} \sum_{\vec{r}\neq (0,0)} 
\frac{e^{i \vec{q}\cdot \vec{r}}}{\vec{r}}
\end{equation}
 
 Without disorder, the total momentum $\vec{K}=\vec{k}_1+\vec{k}_2$ 
is conserved. The two particles states of different momenta are not 
coupled, $H$ becoming block diagonal, each block corresponding to a 
given total momentum $\vec{K}$. 

The dimension of the two particle Hilbert space is given by
\begin{equation}
N_H = \frac{M!}{N! (M-N)!}=\frac{M-1}{2} M 
\end{equation}
for $N=2$ and $M=L^2$. When $L$ is odd, this gives $M$ blocks of 
dimension $(M-1)/2$. When $L$ is even, we have two different 
dimensions of the blocks. 
\begin{equation}
N_H = M_{b_1} M_1 + M_{b_2} M_2
\end{equation}
where $M_1$, $M_2$ are the number of blocks with dimensions $M_{b_1}$, 
$M_{b_2}$. Here, 
\begin{equation}
M_1 = \frac{M\, M_{b_2}-N_H}{M_{b_2}-M_{b_1}}, \qquad 
M_2 = \frac{N_H-M\, M_{b_1}}{M_{b_2}-M_{b_1}} 
\end{equation}
and 
\begin{equation}
M_{b_1} = \frac{M}{2}-2, \qquad M_{b_2} = \frac{M}{2}.  
\end{equation}

\section{The Fermi limit}
\label{sec:U=0}

When $U=W=0$ and with periodic BCs, the states are $N_H$ plane wave 
Slater determinants 
\begin{equation}
a^{\dagger}_{\vec{k}_1}a^{\dagger}_{\vec{k}_2}|0\rangle = 
|\vec{k}_1 \vec{k}_2\rangle.
\end{equation}
$k_x = (2\pi/L) n_x$ and $k_y = (2\pi/L) n_y$ with 
\begin{equation}
n_x \, , \, n_y = \left\{ \begin{array}{l}
0, \pm 1, \ldots , \pm \frac{L-1}{2} \qquad \qquad \,\,  \mbox{for $L$ odd} \\
0, \pm 1, \ldots , \pm \left(\frac{L}{2}-1\right), 
\frac{L}{2} \quad \mbox{for $L$ even}
\end{array} \right.
\end{equation}

The two particle wave functions 
\begin{equation}
\psi \left(\vec{r}_1, \vec{r}_2 \right) = \frac{1}{\sqrt{2}} \left| 
\begin{array}{cc}
\frac{1}{L} e^{i\vec{k}_1 \cdot \vec{r}_1} & 
\frac{1}{L} e^{i\vec{k}_1\cdot \vec{r}_2} \\ \\
\frac{1}{L} e^{i\vec{k}_2 \cdot \vec{r}_1} & 
\frac{1}{L} e^{i\vec{k}_2 \cdot \vec{r}_2}
\end{array} \right| 
\end{equation}
become 
\begin{equation}
\psi \left(\vec{r}_1,\vec{r}_2 \right) = \left( \frac{1}{L} 
e^{i \vec{K}\cdot\vec{R}} \right) \left( i\frac{\sqrt{2}}{L} \sin 
{\vec{k}\cdot\vec{r}} \right)
\end{equation}
after the change of coordinates: 
\begin{eqnarray}
\vec{K} & = & \vec{k}_1 + \vec{k}_2, \qquad 
\vec{k} = \frac{1}{2} \left( \vec{k}_2 - \vec{k}_1 \right) \\
\vec{R} & = & \frac{1}{2} \left(\vec{r}_1 + \vec{r}_2 \right), \qquad 
\vec{r} = \vec{r}_2 - \vec{r}_1
\end{eqnarray}
Without disorder, the motion of the center of mass and the relative 
motion are separable: 
\begin{equation}
\psi\left(\vec{r}_1, \vec{r}_2\right) = 
\phi\left(\vec{R} \right) \chi\left(\vec{r}\right).
\end{equation}
$\phi\left( \vec{R} \right)$ is a plane wave of total momentum $\vec{K}$ 
which describes the propagation of the center of mass on the 
$2d$ torus while $\chi\left(\vec{r}\right)$ describes the 
relative motion. The moments $\langle |\vec{r}|^m \rangle$ of the 
inter-particle 
spacing $\vec{r}$ are given by:
\begin{equation}\label{rmavecont}
\langle |\vec{r}|^m \rangle = \int |\vec{r}|^m p\left( \vec{r} \right) 
d\vec{r}
\end{equation}
where $p(\vec{r}) = |\chi(\vec{r})|^2$ is the inter-particle spacing 
distribution. 

 The one particle energies $\varepsilon (\vec{k})$ can be ordered by 
increasing values. The one particle GS energy $\varepsilon_0$ 
is not degenerate, while the three next excitations 
$\varepsilon_1$, $\varepsilon_2$, $\varepsilon_3$, are four-fold 
degenerate; the fourth excitation $\varepsilon_4$ is ten-fold degenerate; 
etc. 

\begin{eqnarray}\label{1partstates}
\varepsilon_0 & = & -4t \nonumber \\
\varepsilon_1 & = & -2t\left( 1 + \cos \frac{2\pi}{L} \right) \nonumber \\
\varepsilon_2 & = & -4t \, \cos \frac{2\pi}{L} \\ 
\varepsilon_3 & = & -2t \left( 1 + \cos \frac{4\pi}{L} \right) \nonumber \\
\varepsilon_4 & = & -2t \left( \cos \frac{2\pi}{L} + \cos 
\frac{4\pi}{L}\right) \nonumber \\ & \cdots & \nonumber 
\end{eqnarray}

The two particle GS consists of one particle of energy 
$\varepsilon_0$ and of a second particle of energy 
$\varepsilon_1$. Because $\varepsilon_1$ has a 
four-fold degeneracy, the two particle GS energy 
\begin{equation}\label{groundU=0}
E_0 = \varepsilon_0 + \varepsilon_1 = -4t - 2t \left( 1 + 
\cos \frac{2\pi}{L} \right)
\end{equation}
is also four-fold degenerate.

\begin{figure}
\centerline{
\epsfxsize=8cm 
\epsffile{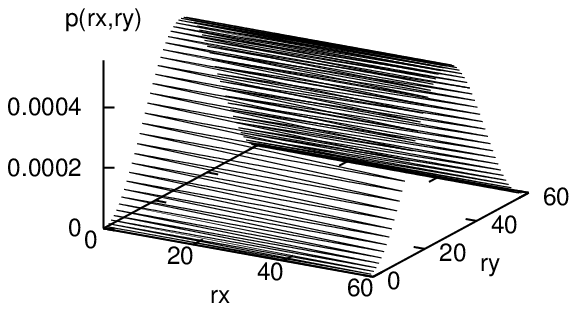}
}
\caption{Inter-particle spacing distribution $p(r_x,r_y)$ for $L=60$, $U=0$
and $\vec{K}=(0,2\pi/L)$.}
\label{fig1}
\end{figure}

 Hereafter, we study the two particle GS of momentum 
${\vec K}=(0,2\pi/L)$. For $\vec{k}_1 = \left( 0, 0\right)$ and 
$\vec{k}_2 = \left( 0, \frac{2\pi}{L} \right)$, the GS wave function 
is given by
\begin{equation}
\psi \left( \vec{r}_1, \vec{r}_2 \right) = \left( \frac{1}{L} 
e^{i\frac{2\pi}{L}R_y} \right) \left( -i\frac{\sqrt{2}}{L} \sin 
\frac{\pi r_y}{L}
\right),
\end{equation}
which yields
\begin{equation}
p\left(r_x,r_y\right) = \frac{2}{L^2} \sin^2 \frac{\pi r_y}{L} 
\end{equation}
\begin{equation}
\label{p(r);U=0}
\left\langle |\vec{r}|^m \right\rangle = \frac{2}{L^2} \sum_{r_x,r_y\neq 0} 
\left( r_x^2 + r_y^2 \right)^{m/2} \sin^2 \frac{\pi r_y}{L}.
\end{equation}
The corresponding inter-particle spacing distribution $p(r_x,r_y)$ is shown in 
Fig. \ref{fig1} for $L=60$.

\begin{figure}
\centerline{
\epsfxsize=8cm 
\epsffile{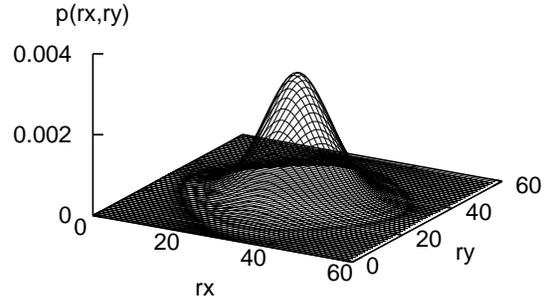}
}
\caption{Inter-particle spacing distribution $p(r_x,r_y)$ for $L=60$, $U/t=5$
and $\vec{K}=(0,2\pi/L)$. }
\label{fig2}
\end{figure}

 When one turns on the interaction $U$, the GS wave function of  
total momentum $\vec{K}=\vec{k}_i+\vec{k}_j$ can be written as
\begin{equation}
\left|\left. \psi_0(U) \right. \right\rangle = \sum_{i<j} 
a_{\vec{k}_i \vec{k}_j}(U) \left| \left. \vec{k}_i \, \vec{k}_j \right. 
\right\rangle
\delta^{\vec{K}}_{\vec{k}_i+\vec{k}_j}.
\end{equation}
The coefficients $a_{\vec{k}_i \vec{k}_j}(U) = \left\langle \vec{k}_i 
\vec{k}_j \left| \psi_0(U) 
\right. \right\rangle$ can be numerically obtained if $L$ is not too large, 
and the inter-particle spacing distribution $p(r_x,r_y)$ given by  
\begin{eqnarray}\label{corrfunctU}
\frac{\sum_{i<j, m<n} a_{\vec{k}_i \vec{k}_j}^* a_{\vec{k}_m \vec{k}_n} 
\delta^{\vec{k}_m+\vec{k}_n}_{\vec{k}_i+\vec{k}_j}
\left[ e^{i(\vec{k}_n-\vec{k}_j)r} - e^{i(\vec{k}_n-\vec{k}_i)r} 
\right]}{L^2} \nonumber
\end{eqnarray}
is shown in Fig. \ref{fig2} for $L=60$, $U/t=5$ and $\vec{K}=(0,2\pi/L)$.
If one particle is located at the site $(0,0)$, one can see how the 
interaction localizes the second one near the site $\vec{r}=(L/2,L/2)$. 
 
\section{The correlated lattice limit}
\label{sec:lattice}

 Before studying in more details the interaction induced localization 
of the inter-particle spacing $\vec{r}$, let us consider the other 
limit $t/U \rightarrow 0$ where the two particle wave functions 
$| \psi \rangle$ are more conveniently written using the operators 
$c_i^{\dagger}$. In this limit, the kinetic part 
\begin{equation}
H_K=-t \sum_{<ij>} c_i^{\dagger} c_j + h.c.
\end{equation}
of $H$ is a small perturbation compared to the Coulomb part and 
the levels can be expanded in powers of $t/U$. When $t/U \rightarrow 0$, 
at the order zero of a $t/U$ expansion, the Coulomb energies $E_0$ and 
$E_1$ of the GS wave function $|\psi_{0}^{(0)}(\vec{K})\rangle$ and of the
two degenerate first excitations $|\psi_{1}^{(0)}(\vec{K})\rangle$ 
and $|\psi_{2}^{(0)}(\vec{K})\rangle$ are given by 
\begin{equation}\label{E0Uinfty}
E_0 = \frac{U}{d_0} =\frac{\sqrt{2}\, U}{L}
\end{equation}
and
\begin{equation}\label{E1Uinfty}
E_1 = \frac{U}{d_1}=\frac{U}{\sqrt{\left(\frac{L}{2}\right)^2 + 
\left( \frac{L}{2}-1\right)^2}}
\end{equation}
if $L$ is even. When $L$ is odd, the lengths $d_0$ and $d_1$ must 
be differently defined:
\begin{equation}\label{dstrongU}
d_0 = \frac{L-1}{\sqrt{2}}
\end{equation}
and 
\begin{equation}\label{distd1}
d_1 = \sqrt{\left(\frac{L-1}{2}\right)^2+\left(\frac{L-1}{2}-1\right)^2}. 
\end{equation}

Hereafter, we assume that $L$ is even and we consider the states 
of total momentum $\vec{K}=(0,2\pi/L)$. $\vec{a}=(L/2,L/2)$.  
$\vec{b}_1=(L/2+1,L/2)$ and $\vec{b}_2=(L/2,L/2+1)$ defining the
locations of three lattice sites which are as far as possible 
from the site $\vec{0}=(0,0)$, the zero order GS wave function
\begin{equation}
| \psi_{0}^{(0)}(\vec{K}) \rangle = \frac{1}{\sqrt{2}L} \sum_{\vec{j}}
\exp^{i\vec{K}.\vec{j}} c^{\dagger}_{\vec{j}}  c^{\dagger}_{\vec{j}+\vec{a}} 
|0 \rangle 
\end{equation}
is directly coupled at the order $t/U$ to the two wave functions of 
energy $E_1$
\begin{equation}
| \psi_{1}^{(0)}(\vec{K}) \rangle =\frac{1}{L} \sum_\vec{j}
\exp^{{i\vec{K}.\vec{j}}} c^{\dagger}_{\vec{j}}  
c^{\dagger}_{\vec{j}+\vec{b}_1} |0 \rangle 
\end{equation}
and
\begin{equation}
|\psi_{2}^{(0)}(\vec{K}) \rangle =\frac{1}{L} \sum_\vec{j}
\exp^{i\vec{K}.\vec{j}} c^{\dagger}_\vec{j}  c^{\dagger}_{\vec{j}+\vec{b}_2} 
|0 \rangle.  
\end{equation}
by   
\begin{equation}
\langle\psi_{1}^{(0)}(\vec{K}) \left| H_K \right|
\psi_{0}^{(0)}(\vec{K}) \rangle =-2 \sqrt{2}\, t
\end{equation}
and 
\begin{equation}
\langle\psi_{2}^{(0)}(\vec{K}) \left | H_K \right|
\psi_{0}^{(0)}(\vec{K}) \rangle =-\sqrt{2}
\left(1+\exp i\frac{2\pi}{L}\right)t
\end{equation}
respectively. At the order $t/U$, the GS wave function  is given by
\begin{equation}
\frac{|\psi_{0}^{(1)}(\vec{K})\rangle}{C^2}=
|\psi^{(0)}_{0}(\vec{K})\rangle
+|\delta\psi_{0}^{(1)}(\vec{K})\rangle
\end{equation}
where $C^{-2}$ is a normalization constant and 
\begin{equation}
|\delta\psi_0^{(1)} (\vec{K})\rangle=
\sum_{\alpha=1}^2 \frac{\langle \psi_{\alpha}^{(0)}(\vec{K})
| H_K |\psi^{(0)}_{0}(\vec{K}) \rangle} {E_0 - E_1}
|\psi^{(0)}_{\alpha}(\vec{K})\rangle.  
\end{equation}
One gets
\begin{equation}
|\delta \psi_{0}^{(1)}(\vec{K})\rangle =  
A|\psi_{1}^{(0)}(\vec{K})\rangle + B |\psi_{2}^{(0)}(\vec{K})\rangle
\end{equation}
where 
\begin{equation}
A=2\sqrt{2}\frac{d_0d_1}{d_0-d_1}\left(\frac{t}{U}\right)
\end{equation}
\begin{equation}
B=\sqrt{2}\frac{d_0d_1}{d_0-d_1}\left(1+\exp{\frac{i2\pi}{L}}\right)
\left(\frac{t}{U}\right)
\end{equation}
and
\begin{equation}
C^{-2}=1+\frac{4d_0^2d_1^2}{(d_0-d_1)^2}\left(3+\cos\frac{2\pi}{L}\right)
\left(\frac{t}{U}\right)^2.
\end{equation}
for $L$ even and $\vec{K}=(0,2\pi/L)$. 
This yields for the fluctuation $\Delta r = \sqrt{\langle r^2\rangle-\langle r
\rangle^2}$ of the
inter-particle spacing:
\begin{equation}
\label{flu}
\Delta r =\frac{t}{U}\, L\, \sqrt{[(L-1)^2+1]\left(3+\cos\frac{2\pi}{L}\right)}
\end{equation}
and for its relative fluctuation  $u_r =\Delta r/\langle r\rangle$
\begin{equation}
\label{flucurL}
u_r = \frac{t}{U}\, \sqrt {2 [(L-1)^2+1] \left(3+\cos\frac{2\pi}{L}\right)}\, .
\end{equation} 

\begin{figure}
\epsfxsize=8cm {\centerline{\epsffile{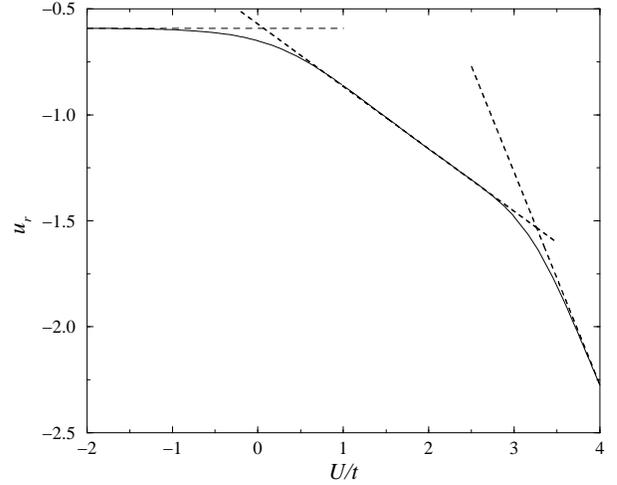}}}
\caption
{
Relative fluctuation $u_r$ as a function of $U$ 
for $L=20$ and $\vec{K}=(0,2\pi/L)$. The log-log plot shows the three 
regimes characterizing the ground state. The three dashed lines 
correspond to $u_r= 0.25$, $u_r \propto (t/U)^{0.31}$ and the
$t/U$ perturbative behavior given by Eq. (\ref{flucurL}). 
}
\label{fig3}
\end{figure}

 This correlated lattice behavior is shown in Fig. \ref{fig3} for
$L=20$ when $U/t > 1000$.

\section{The three regimes of coupling}
\label{sec:Wigner}

 The behavior of the relative fluctuation $u_r=\Delta r /<r>$ 
is shown in Fig. \ref{fig3} for $L=20$
and $\vec{K}=(0,2\pi/L)$. The crossover from independent particle towards 
strongly correlated motion in a finite size lattice exhibits three 
regimes: 

\begin{itemize}

\item A weak $U$ Fermi regime where $u_r$ is large and almost  
independent of $U$. The fluctuation $\Delta r$ is of the order 
of the expectation value $\langle r\rangle$, $r$ being broadly distributed.

\item An intermediate correlated Wigner regime where $u_r$ exhibits a 
weak $(t/U)^{\alpha}$ decay, where $\alpha\approx 0.31$.  $\Delta r$ 
becomes small compared to $\langle r\rangle$, the distribution of $r$ becoming
narrower. 

\item A large $U$ strongly correlated lattice regime where $u_r$ exhibits the  
strong $t/U$ decay predicted by the perturbative expansion 
(Eq. (\ref{flucurL})), the distribution of $r$ being extremely narrow. 

\end{itemize}

 The value $U^*(L)/t$ characterizing the crossover between the intermediate 
Wigner regime and the strongly correlated lattice regime is consistent 
with the condition $\Delta r < 1$. When $U \approx U^*(L)$, the lattice 
strongly reduces the degrees of freedom of the relative motion, and 
lattice expansions in powers of $t/U$ become valid. 
We study in the next section the Fermi-Wigner crossover which 
occurs before this strongly correlated lattice regime.

\section{Scaling theory of the Fermi-Wigner crossover}
\label{sec:scaling}

\begin{figure}
\epsfxsize=8cm {\centerline{\epsffile{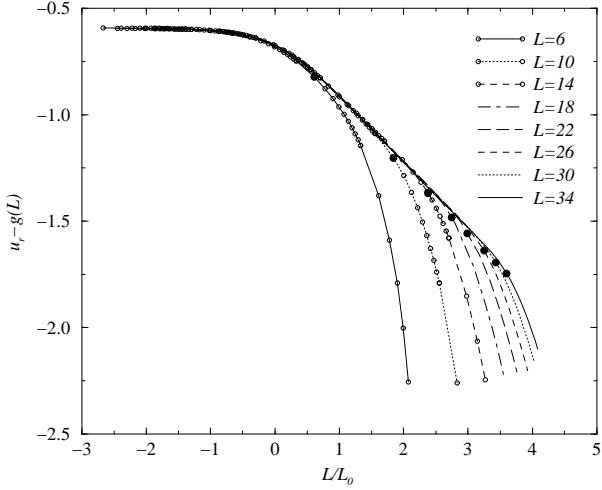}}}
\caption
{
Relative fluctuation $u_r - g(L)$ where 
$g(L)= 0.9/(L+1)^2$ as a function of 
$L/L_0 \approx UL/({\sqrt8}\pi^2 t)$ for even 
$L$. The log-log plot shows the universal scaling function $F(L/L_0)$ 
for $U<U^*(L)$, and the non universal lattice regime for 
$U>U^*(L)$. The filled circles approximately 
give the thresholds $U^*(L)$.
}
\label{fig4}
\end{figure}

\begin{figure}
\epsfxsize=9cm {\centerline{\epsffile{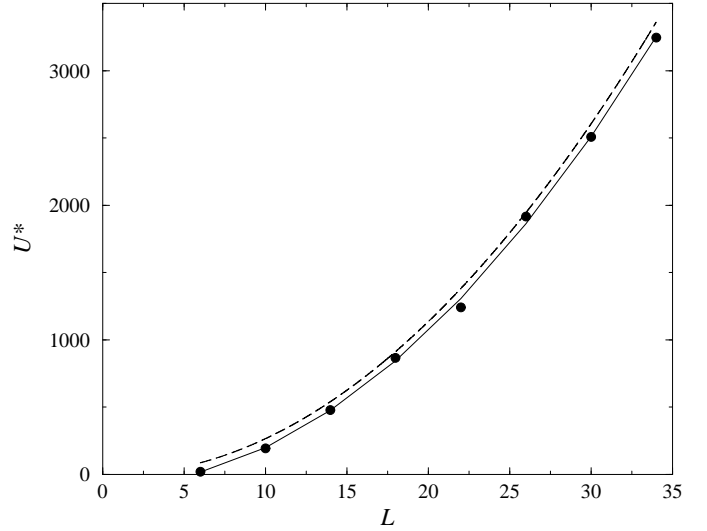}}}
\caption
{
Circles: Interaction thresholds $U^*(L)$ giving the end of the 
universal  scaling regime and the beggining of the strongly 
correlated lattice regime, extracted from Fig. \ref{fig4}. The dashed
line corresponds to the condition $\Delta r (U) = 2/3$ where $\Delta r$ 
is given by Eq. (\ref{flu}).
}
\label{fig5}
\end{figure}

 For $U<U^*(L)$, we define the scale $L_0$ associated to the 
interaction strength $U/t$ such that a dimensionless observable 
as $u_r$ becomes a universal function of the dimensionless ratio 
$L/L_0$, up to certain finite size corrections which we will estimate. 
The argument for defining the scaling length $L_0$ can be presented 
as follows. Let us consider the system at $U=0$ where the first particle 
occupies the state $\varepsilon_0=-4t$. Due to Pauli principle, 
adding a second particle requires an extra kinetic energy:  
\begin{equation}
\Delta E(U=0)=\varepsilon_1-\varepsilon_0 = 2t(1-\cos(2\pi/L))
\end{equation}  
In the other limit $t \rightarrow 0$, adding a second particle requires 
an extra Coulomb energy: 
\begin{equation}
\Delta E (t=0) = \frac{U}{d_0}
\end{equation}

 The length $L_0$ characterizes the scale at which $\Delta E (U=0)$ 
coming from Pauli exclusion principle is equal to $\Delta E (t=0)$ 
coming from Coulomb repulsion. When $L$ is large enough, $L_0$ can 
be approximated by: 
\begin{equation}
L_0 = \sqrt{8} \pi^2 \left(\frac{t}{U}\right),
\end{equation}
but when $L$ is small, $L_0$ should be more precisely determined 
by solving the equation $\Delta E (U=0)=\Delta E (t=0)$, and 
without neglecting the even-odd effects in the definition of $d_0$.
For scales $L < L_0$, the GS wave function has mainly to minimize the 
kinetic energy while the minimization of the Coulomb energy becomes 
more important for the scales $L>L_0$. The Fermi-Wigner crossover is 
expected at $L/L_0=1$. 

 One assumes for the dimensionless ratio $u_r$ the usual finite size scaling 
ansatz \cite{scaling}: 
\begin{equation}
u_r(L,U,t) = F (\frac{L}{L_0}) + g (L),
\label{ansatz}
\end{equation}
the finite size correction $g(L) \rightarrow 0$ as 
$L \rightarrow \infty$. $g(L)$ can be easily evaluated when 
$U/t \rightarrow 0$ ($L_0 \rightarrow\infty$), the ansatz 
becoming  
\begin{equation}
u_r(L,U=0,t) = F(0) + g(L)
\end{equation} 
This is a problem without interaction where one can use 
Eq. (\ref{p(r);U=0}), for eventually obtaining $F(0)=0.2543$ and 
\begin{equation}
g(L) \approx \left\{ \begin{array}{cl} 
0.9/\left( L + 1 \right)^2 & \mbox{for $L$ even} \\
-0.4/ \left( L - 1 \right)^2 & \mbox{for $L$ odd}
\end{array} \right.
\end{equation}
The finite size correction $g(L)$ has a sign which 
depends on the parity of $L$ and disappears as the inverse 
of the total number of lattice sites. 

Fig. \ref{fig4} gives $F(L/L_0)=u_r - g(L)$ for even values of $L$.
One can see the universal scaling behavior with a Fermi Wigner 
crossover at $L=L_0$ up to a value $U^*(L)$ denoted by the filled 
circles where scaling breaks down and the correlated lattice regime 
begins. The behavior of $U^*(L)$ is shown in Fig. \ref{fig5},
where one can see that the values $U^*(L)$ extracted from 
Fig. \ref{fig4} are precisely given by the condition
$\Delta r \approx 2/3$, $\Delta r$ being given by Eq. (\ref{flu}). 
Below $U^*(L)$, $u_r-g(L)$ is a one parameter function of $L/L_0 
\propto LU/t$, and the lattice effects are irrelevant. The lattice 
regime occurs only above a large interaction $U^*/t \propto L^2$ 
when $L$ is large. 

For $N$ spinless fermions in a $L \times L$ square lattice, the Coulomb 
energy to Fermi energy ratio becomes \cite{bwp}: 
\begin{equation}
r_s=\frac{U}{2t}\frac{1}{\sqrt {\pi \nu}}
\end{equation}
for a density $\nu=N/L^2$. As it must be, the dimensionless ratios 
$L/L_0$ and $r_s$ are proportional, the advantage of using $L/L_0$ 
instead of $r_s$ being that the crossover occurs when $L/L_0 = 1$, 
and not at $r_s =1$. In the strongly correlated lattice regime, $u_r$ 
becomes a function of $Lt/U$, and not of $r_s \propto LU/t$. 

\begin{figure}
\epsfxsize=8cm {\centerline{\epsffile{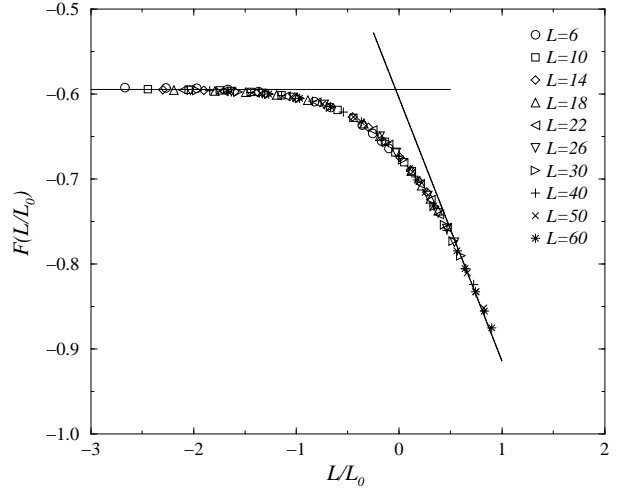}}}
\caption{
Fermi-Wigner crossover for the relative fluctuation $u_r$, 
described by the universal scaling function $F(L/L_0) = u_r - g(l)$ 
in a log-log plot. Values calculated from even values of $L$ with the 
finite size correction $g(L)= 0.9/(L+1)^2$.  
The solid lines correspond to $F(x)=0.2543$ and $F(x) \propto  
x^{-0.31}$ respectively.
}
\label{fig6}
\end{figure}

\begin{figure}
\epsfxsize=8cm {\centerline{\epsffile{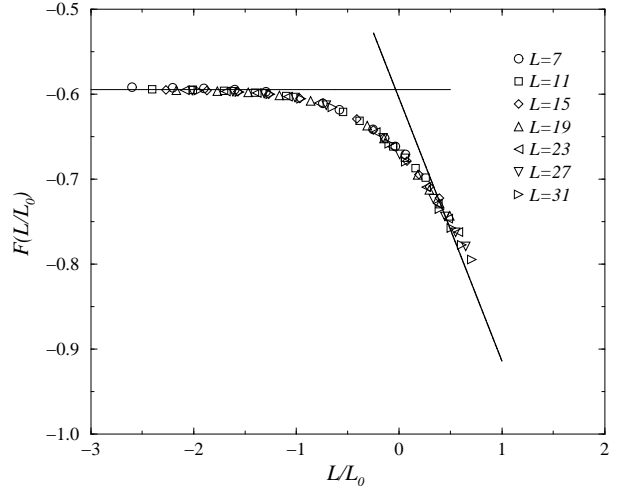}}}
\caption
{
Fermi-Wigner crossover for the relative fluctuation $u_r$, 
described by the universal scaling function $F(L/L_0) = u_r - g(L)$ 
in a log-log plot. Values calculated for odd values of $L$ with the finite 
size correction $g(L)= -0.4 /(L-1)^2$. 
}
\label{fig7}
\end{figure}

 The Fermi-Wigner universal crossover is illustrated with more details in 
Fig. \ref{fig6} where we have only taken
even values of $L$. One can see that the finite size correction 
$g(L) = 0.9 /(L+1)^2$ allows us to plot all the values $u_r-g(L)$ 
calculated for $U<U^*(L)$ onto the universal scaling curve 
$F(L/L_0)$ when $L=6, 10, \dots ,60$. Fig. \ref{fig7} gives
the same universal curve $F(L/L_0)$ for the odd values of $L$, once the 
finite size correction $g(L)=-0.4/(L+1)^2$ have been subtracted to $u_r$. 
The universal function $F(x)$ is close to its value $F(0)= 0.2543$ when 
$x \leq 1$ and behaves as $x^{-\alpha}$ with $\alpha \approx 0.31$  when 
$x \geq 1$. 

The value $\alpha \approx 0.31$ depends on the exact form of the pairwise 
repulsion $U/|\vec{r}|$ around $\vec{r}=(L/2,L/2)$. Due to our definition 
of the inter-particle spacing $r$, $U(r)$ is non analytic around 
$\vec{r}=(L/2,L/2)$, making the study a little bit involved. If one modifies 
the pairwise repulsion such that $U(r)$ becomes analytic around $(L/2,L/2)$, 
taking for instance
\begin{equation} 
U(\vec{r})
= 
\frac {U} 
{
\frac{L}{\pi} 
\sqrt{\sin ^2 (\frac{r_x \pi}{L}) + \sin ^2 (\frac{r_y \pi}{L})} 
}
\end{equation} 
instead of $U/|\vec{r}|$, one can show \cite{nemeth} that the equation 
for the relative motion becomes identical to the single particle motion 
in a $2d$ harmonic potential when $U$ is large enough, to eventually 
obtain a slightly different exponent $\alpha={1/4}$ in the continuous 
limit ($L\rightarrow \infty$).


\section{Effects of a weak random substrate}
\label{sec:disorder}

\begin{figure}
\epsfxsize=8cm {\centerline{\epsffile{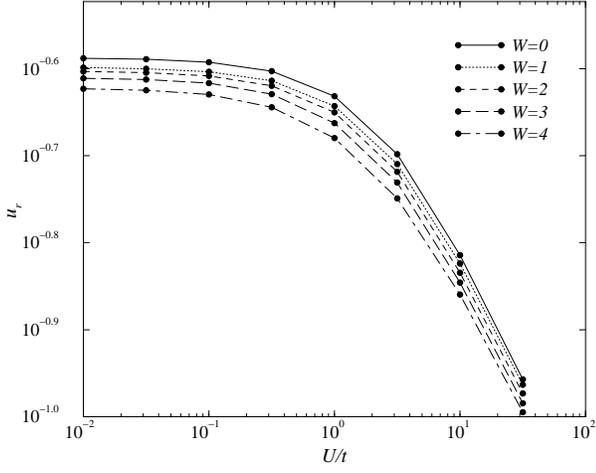}}}
\caption
{
Ensemble average relative fluctuation $\langle u_r\rangle$ as a function of
$U/t$ for $L=14$ and $W=0,1,2,3,4$ in a log-log plot. 
}
\label{fig8}
\end{figure}
\begin{figure}
\centerline{
\epsfxsize=9cm 
\epsffile{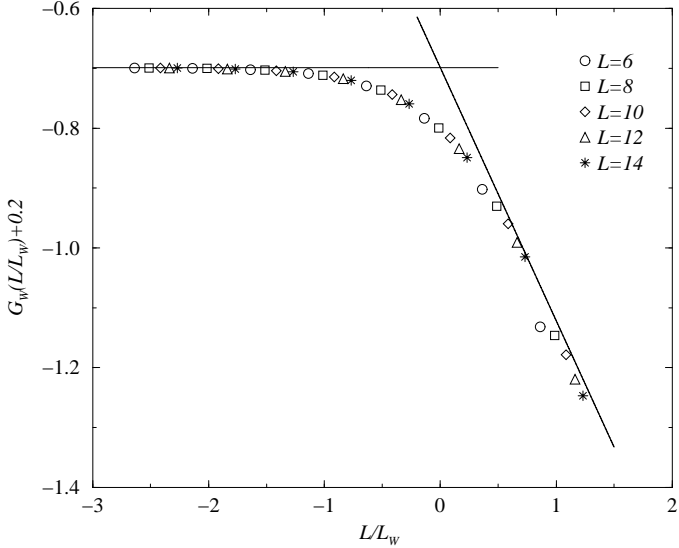}
}
\caption
{
Scaling function $G_W(L/L_W)+C$ for $W=2$ and $6 \leq L \leq 14$.
An arbitrary constant $C=0.2$ has been taken for the log-log plot. 
}
\label{fig9}
\end{figure} 
\begin{figure}
\centerline{
\epsfxsize=9cm 
\epsffile{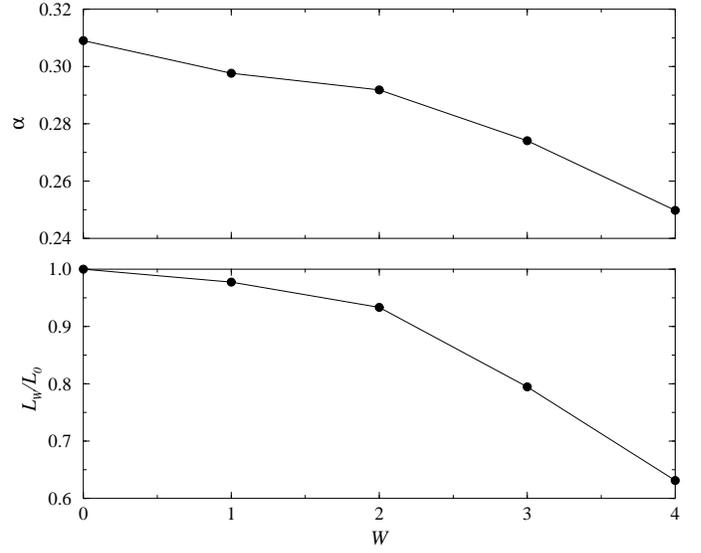}
}
\caption
{ As a function of $W$, exponent $\alpha (W)$ characterizing the 
the scaling function $G_W(x) \propto x^{-\alpha(W)}$ when $x>1$ 
(upper figure) and reduction factor $L_W/L_0$ for the 
scaling length $L_W$ (lower figure).
}
\label{fig10}
\end{figure}

  We now study the effect of random site potentials $v_i$ upon the 
Fermi-Wigner crossover. We restrict our study to the limit of weak 
disorder, the values of $W$ being large enough to yield one particle 
diffusive dynamics in the non interacting system, but too small 
to yield one particle Anderson localization. In the presence of disorder, 
the momenta $\vec{K}$ are not conserved, and we need to diagonalize the 
total Hamiltonian $H$ given by Eq. (\ref{siteH}) in the site basis 
$c_i^{\dagger}c_j^{\dagger}|0\rangle$. Moreover, the relative fluctuations
$u_r$ are now random variables, and the ensemble average values
$\langle u_r\rangle$ will be considered.

 Figure \ref{fig8} shows for $L=14$ the difference between the
$u_r$ obtained for $W=0$ and the $\langle u_r\rangle$ calculated for
$1\leq W \leq4$. One can see that the disorder reduces the relative fluctuations
$\langle u_r\rangle$, shifts the Fermi-Wigner crossover to weaker
interactions, and reduces the exponent $\alpha$ characteristic of the
$U^{-\alpha}$ decay for intermediate interactions.

 For a weak value of $W$ and small enough values of $U/t$ to avoid 
the correlated lattice regime, we assume the generalized 
scaling ansatz : 
\begin{equation} 
\langle u_r(L,U,t,W)\rangle=F_W(\frac{L}{L_W(U,t,W)})+g_W(L).
\label{ansatzD}
\end{equation} 
Since  
\begin{equation} 
\langle u_r(L,U=0,t,W)\rangle=F_W(0)+g_W(L),
\label{U=0}
\end{equation} 
which implies that 
\begin{equation}
G_W=\langle u_r(L,U,t,W)\rangle-\langle u_r(L,U=0,t,W)\rangle
\end{equation} 
must only depend on the ratio $L/L_W$.  The scaling length $L_W$ is 
defined in such a way that the Fermi-Wigner crossover remains 
at $L/L_W=1$. As shown in Fig. \ref{fig9}, all the data
calculated for $L=8,10,12,14$ can be mapped onto the same scaling 
curve $G_W(L/L_W)=F_W(L/L_W)-F_W(0)$. We just show the curve 
$G_{W=2}(L/L_W)$, the curves obtained for $W=1,2,3,4$ 
being also consistent with the ansatz (\ref{ansatzD}) when $U$ and 
$L$ are varied in the same range.  $G_W(x) \approx 0$ 
for $ x < 1$ while $G_W(x) \approx x^{-\alpha(W)}$ for $x >1$, the 
exponent $\alpha(W)$ being given in upper Fig. \ref{fig10}.
To obtain that the Fermi-Wigner crossover remains at $L/L_W=1$, 
one needs to take the scaling lengths $L_W$, the reduction factors 
$L_W/L_0$ being given  in the lower Fig. \ref{fig10}.

\begin{figure}
\centerline{
\epsfxsize=9cm 
\epsffile{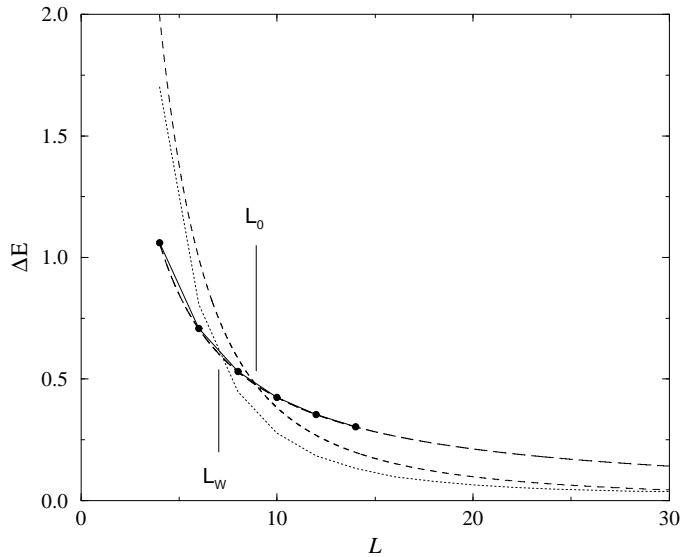}
}
\caption
{ As a function of $L$, curves $\Delta E(U=0,t=1,W=0)$ (dashed line) 
and $\Delta E(U=3,t=0,W=0)$ (long dashed line) crossing at $L=L_0$, 
and ensemble average curves $\langle \Delta E(U=0,t=1,W=2)\rangle$
(dotted line) $\langle \Delta E(U=3,t=0,W=2)\rangle$ (filled circle) crossing
at $L=L_W$.
$L_0 (t/U) \approx 8.9$ and $L_{W=2} (t/U) \approx 7.1$. 
}
\label{fig11}
\end{figure}

 The reduction of the crossover scale $L_W$ by the random potentials 
can be qualitatively explained if we revisit the argument which has 
allowed us to define $L_0$ without disorder: $L=L_W$ when the energies 
$\Delta E_W$ which are necessary for adding a second particle in the 
one particle system are the same in the two limits $U=0$ and $t=0$. 
On one hand, when $U=0$, the first one particle excitation has a 
fourfold degeneracy which is removed by disorder. This gives a reduction 
of the energy cost $\langle \Delta E_W (U=0)\rangle$ necessary in the Fermi
limit. On the other hand, the energy cost $\langle \Delta E_W(t=0)\rangle$
necessary in the
Coulomb limit remains almost unchanged if $W$ is weak. This gives a shift 
of the Fermi-Wigner crossover to a lower scale $L_W < L_0$ in the limit 
of a weak disorder. In Fig. \ref{fig11}, we have plotted as a
function of $L$ for $W=0$ and $W=2$ the disorder averaged energies 
$\langle \Delta E_W (U=0)\rangle$ and $\langle \Delta E_W(t=0)\rangle$.
One can see indeed that $\langle \Delta E_W (U=0)\rangle$ is decreased by the
disorder, while $\langle \Delta E_W(t=0)\rangle$
remains essentially unchanged. The curves cross at $L=L_0\approx 8.9$ when 
$W=0$ and $L=L_W \approx 7.1$ when $W=2$. The obtained ratio $L_{W=2}/L_0 
\approx 0.8$ is indeed smaller than $1$, though somewhat below the value 
$L_{W=2}/L_0 \approx 0.9$ given in Fig. \ref{fig10}.

 This quantitative disagreement may be due the distributions of the 
considered random variables. We have considered only the ensemble
average behaviors, which may not give the typical scaling behaviors, 
both for the relative fluctuations and for the energies. When $U=0$, 
it is indeed well known that the level spacing distributions are not 
normally distributed, but depend on the nature of the one particle dynamics. 
In the bulk of the spectrum, one has typically a Poisson distribution 
if the dynamics is ballistic and non chaotic, a Wigner-Dyson distribution 
if the dynamics is diffusive, a  Poisson distribution again when one 
has Anderson localization. Near the one particle spectrum edges, the 
distribution can differ from the bulk distribution. This is why 
an argument based on disorder average quantities can only qualitatively 
give the disorder induced reduction of the scaling length. 

In conclusion, let us underline that an Hamiltonian with three parameters, 
$U$, $t$ and $W$, has many different limits. This section is restricted to 
the study of the competition between the interaction $U$ and the 
kinetic energy $t$ when the disorder $W$ remains a weak perturbation, in 
the limit $L<L_1$, $L_1$ denoting the one particle localization length. In 
this limit, we have shown that $\langle u_r\rangle$ is still given by a function
$F_W$ of the dimensionless ratio $L/L_W$ and that the scaling length 
$L_W$ decreases as a function of $W$. This is in agreement with the general 
idea that a random substrate favors charge crystallization \cite{bwp,chui}, 
the correlated motion persisting to larger densities (weaker values of 
$r_s$). But for larger $W$, the issue is to describe the crossover from a 
Fermi glass of Anderson localized states towards a Wigner glass. In this 
strongly disordered limit, the $u_r$ will have large sample to sample 
fluctuations, and one cannot rule out that the average $\langle u_r\rangle$ will
finish
to be meaningless or that the one parameter scaling function $F_W(L/L_W)$ 
could become a more complicated two parameter scaling function of $L/L_0$ 
and $L/L_1$. The study of those glassy behaviors is left for another study.

\begin{acknowledgement}
CONACyT-Mexico and CEA have supported M. Mart\'\i nez during the time 
this research was performed in Saclay. We thank Z. \'A. N\'emeth for useful 
discussions. 
\end{acknowledgement}


\begin{thebibliography}{99} 

\bibitem{Egger} R.\ Egger, W.\ H\"ausler, C.H.\ Mak and 
H.\ Grabert, Phys.\ Rev.\ Lett.\ {\bf 82}, 3320 (1999).

\bibitem{yannouleas1} C.\ Yannouleas and U.\ Landman, 
Phys.\ Rev.\ Lett.\ {\bf 82}, 5325 (1999).

\bibitem{yannouleas2} C.\ Yannouleas and U.\ Landman, 
Phys.\ Rev.\ Lett.\ {\bf 86}, 1726 (2000).

\bibitem{yannouleas3} C.\ Yannouleas and U.\ Landman, 
Phys.\ Rev.\ {\bf B 61}, 15895 (2000).

\bibitem{peeters} A.\ Matulis and F. M. \ Peeters, cond-mat/0005299.
 
\bibitem{hawrylak} J.\ Kyriadidis, M.\ Pioro-Ladriere, M.\ Cioga, A.S.\ 
Sachrajda and P.\ Hawrylak, cond-mat/0111543.

\bibitem{filinov} A.V.\ Filinov, M. Bonitz and Yu.E.\ Lozovik, 
Phys.\ Rev.\ Lett.\ {\bf B 86}, 3851 (2001).

\bibitem{dot} L.\ Jacak, P.\ Hawrylak and A.\ Wojs, 
Quantum dots, (Springer-Verlag, Berlin Heidelberg, 1998).

\bibitem{schweigert} I.V.\ Schweigert, V.A.\ Schweigert and F.M.\ Peeters, 
Phys.\ Rev.\ Lett.\ {\bf 84}, 4381 (2000).

\bibitem{scaling} M.E.\ Fisher and M.N.\ Barber, Phys.\ Rev.\ Lett.\ 
{\bf28}, 1516 (1972); J.-L.\ Pichard and G.\ Sarma, J.\ Phys.\ C:\ 
Solid State Physics\ {\bf14}, L127 and L617 (1981) and refs. therein.

\bibitem{bwp} G.\ Benenti, X.\ Waintal, and J.-L.\ Pichard, 
Phys.\ Rev.\ Lett. {\bf 83}, 1826 (1999).

\bibitem{nemeth} Z. \'A.\ N\'emeth, unpublished.
 
\bibitem{chui} S.T.\ Chui and B.\ Tanatar, Phys.\ Rev.\ Lett. {\bf 74}, 
458 (1995).

\end{thebibliography}
\end{document}